\documentclass[12pt]{article}
\usepackage{amsfonts,amssymb}
\usepackage{amsmath}

\def\hook{\hbox{\vrule height0pt width4pt depth0.3pt
\vrule height7pt width0.3pt depth0.3pt \vrule height0pt width2pt
depth0pt} }
\newcommand{\sect}[1]{\setcounter{equation}{0}\bigskip\medskip
\section{#1}\smallskip}
\newcommand{\subsect}[1]{\medskip\subsection{#1}\smallskip}

\newtheorem{THEOREM}{Theorem}[section]  
\newenvironment{theorem}{\begin{THEOREM} \hspace{-.85em} {\bf :} 
}%
                        {\end{THEOREM}}
\newtheorem{LEMMA}[THEOREM]{Lemma}
\newenvironment{lemma}{\begin{LEMMA} \hspace{-.85em} {\bf :} }%
                      {\end{LEMMA}}
\newtheorem{COROLLARY}[THEOREM]{Corollary}
\newenvironment{corollary}{\begin{COROLLARY} \hspace{-.85em} {\bf 
:} }%
                          {\end{COROLLARY}}
\newtheorem{PROPOSITION}[THEOREM]{Proposition}
\newenvironment{proposition}{\begin{PROPOSITION} \hspace{-.85em} 
{\bf :} }%
                            {\end{PROPOSITION}}
\newtheorem{DEFINITION}[THEOREM]{Definition}
\newenvironment{definition}{\begin{DEFINITION} \hspace{-.85em} {\bf 
:} \rm}%
                            {\end{DEFINITION}}
\newtheorem{EXAMPLE}[THEOREM]{Example}
\newenvironment{example}{\begin{EXAMPLE} \hspace{-.85em} {\bf :} 
\rm}%
                            {\end{EXAMPLE}}
\newtheorem{CONJECTURE}[THEOREM]{Conjecture}
\newenvironment{conjecture}{\begin{CONJECTURE} \hspace{-.85em} 
{\bf :} \rm}%
                            {\end{CONJECTURE}}
\newtheorem{PROBLEM}[THEOREM]{Problem}
\newenvironment{problem}{\begin{PROBLEM} \hspace{-.85em} {\bf :} 
\rm}%
                            {\end{PROBLEM}}
\newtheorem{REMARK}[THEOREM]{Remark}
\newenvironment{remark}{\begin{REMARK} \hspace{-.85em} {\bf :} 
\rm}%
                            {\end{REMARK}}
\newtheorem{CONCLUSION}[THEOREM]{Conclusion}
\newenvironment{conclusion}{\begin{CONCLUSION} \hspace{-.85em} {\bf :} 
\rm}%
                            {\end{CONCLUSION}}
\newcommand{\thm}{\begin{theorem}}
\newcommand{\lem}{\begin{lemma}}
\newcommand{\pro}{\begin{proposition}}
\newcommand{\dfn}{\begin{definition}}
\newcommand{\rem}{\begin{remark}}
\newcommand{\con}{\begin{conclusion}}
\newcommand{\xam}{\begin{example}}
\newcommand{\cnj}{\begin{conjecture}}
\newcommand{\prb}{\begin{problem}}
\newcommand{\cor}{\begin{corollary}}

\newcommand{\ethm}{\end{theorem}}
\newcommand{\elem}{\end{lemma}}
\newcommand{\epro}{\end{proposition}}
\newcommand{\edfn}{\bbox\end{definition}}
\newcommand{\erem}{\bbox\end{remark}}
\newcommand{\econ}{\bbox\end{conclusion}}
\newcommand{\exam}{\bbox\end{example}}
\newcommand{\ecnj}{\bbox\end{conjecture}}
\newcommand{\eprb}{\bbox\end{problem}}
\newcommand{\ecor}{\end{corollary}}

\newcommand{\beqn}{\begin{equation}}
\newcommand{\eeqn}{\end{equation}}

\newcommand{\bbox}{\vrule height7pt width4pt depth1pt}

\def\br{\begin{eqnarray}}
\def\er{\end{eqnarray}}
\def\brn{\begin{eqnarray*}}
\def\ern{\end{eqnarray*}}
\def\er{\end{eqnarray}}
\def\eqq{&\!\!\!\!\!=\!\!\!\!\!&}
\def\eee{&\!\!\!\!\!\!\!\!\!\!&}
\def\vt{\vartheta}

\def\a{\alpha}
\def\b{\beta}

\def\R{\overline {R}}
\def\r{\overline {r}}
\def\v{\overline {v}}
\def\e{\overline {e}}
\def\p{\overline {\psi}}
\def\eqq{&\!\!\!\!\!=\!\!\!\!\!&}
\def\eee{&\!\!\!\!\!\!\!\!\!\!&}
\def\dps{\dot{\overline {\psi}}}
\def\ddps{\ddot{\overline {\psi}}}
\def\dddps{\dot{\ddot{\overline {\psi}}}}

\title{\bf{Equation of motion in a scalar model of gravity}}
\author{
\thanks {\quad   email kaniel@math.huji.ac.il}
\small {\rm Shmuel Kaniel and
\thanks {\quad   email itin@math.huji.ac.il}
\small {\rm Yakov Itin}}\\
\small {\rm Institute of Mathematics}\\
\small {\rm Hebrew University of Jerusalem}\\
\small {\rm Givat Ram, Jerusalem 91904, Israel}\\
}
\begin{document}  

\newcommand{\bi}[1]{\bibitem{#1}}
\date{\today}

\maketitle
\begin{abstract}
\def\sqr#1#2{{\vcenter{\hrule height.#2pt\hbox{\vrule width.#2pt
height#1pt \kern#1pt \vrule width.#2pt}\hrule height.#2pt}}}
\def\square{\mathchoice\sqr64\sqr64\sqr{4.2}3\sqr{3.0}3 \ \!} 
\def\hatsquare { {\hat\sqcap\!\!\!\!\sqcup} }
A scalar model of gravity is considered. We propose Lorentz invariant  field equation 
$\square f=k\eta_{ab}f_{,a}f_{,b}$. The aim of this model is to get, approximately, Newton's law of gravity. It is shown that $f=-\frac 1k\ln(1-k\frac mr)$ is the unique spherical symmetric static solution of the field equation. $f$ is taken to be the field of a particle at the origin, having the mass $m$. The field of a particle moving with a constant velocity is taken to be the appropriate Lorentz transformation of $f$. The field $F$ of $N$ particles moving on trajectories $\overline{\psi_j(t)}$ is taken to be, to first order, the superposition of the fields of the particles, where the instantaneous Lorentz transformation of the fields pertaining to the $j$-th particle is  $\overline{\dot\psi_j(t)}$. When this field is inserted to the field equation the outcome is singular at $(\overline{\psi_j(t)},t)$. The singular terms of the l.h.s. and of the r.h.s. are both $O(R^{-2})$. The only way to reduce the singularity in the field equation is by postulating Newton's law of force.\\
It is hoped that this model will be generalized to system of equations that are covariant under general diffeomorphism.
\end{abstract}
\sect{Introduction}
Classical field theory postulates the world to be a 4D-differential manifold $M$. The evolution in $M$ is represented by field equations. The most fundamental are: the scalar wave equation, Maxwell and Einstein field equations. To these one may add Newton equation for the gravitational potential. It is not an evolution equation, but a field equation for a section $t=\text{const}$. For the field equations above particles are postulated to be the singularities of the field. Newton and Maxwell equations are linear. Therefore, there is no way to formulate, only by the field equations, the ``force'' that a particle exerts on other particles. Thus additional laws were included: Newton's law of gravitation, Coulomb and Lorentz laws in electrodynamics.\\
On the other hand, Einstein field equation for gravitation is non linear. This allows, in theory, to derive equations of motion from the field equations. Einstein succeeded to carry out this program. The original version of general relativity (GR) employed an additional postulate: The Geodesic Postulate. Thus, he obtained a covariant generalization of the classical equation of motion. The Geodesic Postulate is an intrinsic postulate involving only the geometry of $M$. By comparison, Newton, Coulomb and Lorentz laws are extraneous laws of force. \\
Later, Einstein- Infeld-Hoffman \cite{EIH} and Fock \cite{Fo} derived, directly, the equations of motion from the field equations. The derivation is rather formal (i.e. the first terms of a  series). Moreover, it is not known how the motion of the particles can be embedded in a field satisfying Einstein field equations. Recently Damour \cite{Dam} computed the equations of motion to great accuracy. Still, the derivation is formal. \\
In this paper we study  a Lorentz-invariant scalar model of gravity. This is  a non-linear (quadratic) Lorentz-invariant generalization of the Newtonian scalar theory of gravity. We derive by this model the equations of particle motion and Newton's law of force from the field equation.\\
Recently, Watt and Misner \cite{WM} considered a scalar model of gravity (It is interesting to note that the metric that they obtained is, also, the metric obtained by Kaniel and Itin \cite{K-I} i.e. Yilmaz-Rosen metric.) The motive of \cite{WM} was to facilitate numerical computations of gravitational waves. We hope that the model established in this paper will be of case also for analytical reasoning. It may serve as guideline for studying the equations of motion, where the field equation is a diffeomorphic covariant system.  
We propose to apply the method exhibited in this paper to general  
 quasi-linear field equations.
The computations above lead to a novel algorithm for the derivation of equations of motion from it.
\begin{itemize}
\item[\bf{1}] Compute a static, spherically symmetric solution of the field equation. It will be singular at the origin. This will be taken to be the field generated by a single particle.
\item[\bf{2}] Move the solution on a trajectory $ \overline{\psi(t)}$ and apply the instantaneous Lorentz transformation based on $\dot{\overline{\psi(t)}}$.
\item[\bf{3}] Take the field generated by $n$ particles to be the superposition of the fields generated by the  single particles. 
\item[\bf{4}] Compute the leading part of the equation. Hopefully, only terms that involves $\ddps$ will be dominant.
\item[\bf{5}]Compute the ``force'' between the particles by the quadratic part of the equation.
\end{itemize}
This algorithm should result, in most cases, in the verification of Newton gravitation law.\\
The method presented in this paper may enable the embedding the trajectories satisfying the equations of motion in a field satisfying the field equations.

\sect{Non-interacting particles} 
We start with the linear d'Alembertian field equation. This equation is a Lorentz-invariant extension of Laplace equation pertaining to the Newtonian gravity.\\
The Newtonian theory of gravity is formulated as a field theory via the scalar potential $f$, which is subject to Laplace equation
\begin{equation}\label{2-1}
\triangle f=0.
\end{equation}
The unique spherical symmetric and asymptotically zero solution of this equation is
\begin{equation}\label{2-2}
f=\frac mR,
\end{equation}
where $R=\sqrt{(\r-\r_0)^2}$,  $m$ and $\r_0$ are arbitrary constants of the integration. This solution is singular,  so it represents a field of a pointwise particle located in the point   $\r=\r_0$. The parameter $m$, taken to be positive, represents the mass of the particle. \\
Laplace equation is not Lorentz-invariant. Thus it should be generalized to the wave field equation 
\begin{equation}\label{2-3}
\square  \ f=0.
\end{equation}
The time independent spherical symmetric asymptotically zero solution of this equation is again 
\begin{equation}\label{2-4}
f=\frac mR,
\end{equation}
where
\begin{equation}\label{2-5}
\R=(\r - \r_0)-\v \Big(\a<\v,(\r-\r_0)>+\b t\Big),
\end{equation}
provided that (\ref{2-5} is a Lorentz transformation ($\a$ and $\b$ are functions of $v$).\\
 We reaffirm this fact in order to establish the notation.\\
This solution describes free motion of a particle.
Indeed, the singularity point satisfies 
\begin{equation}\label{2-6}
 \r - \r_0=\v \Big(\a<\v,(\r-\r_0)>+\b t\Big).
\end{equation}
So
\begin{equation}\label{2-7}
<\v,(\r-\r_0)>=v^2\Big(\a<\v,(\r-\r_0)>+\b t\Big).
\end{equation}
Or
\begin{equation}\label{2-8}
<\v,(\r-\r_0)>=\frac{v^2\b t}{1-\a v^2}.
\end{equation}
Substituting this relation in (\ref{2-6}) we obtain the equation motion of  the singularity as
\begin{equation}\label{2-9}
\r-\r_0=\frac{\b}{1-\a v^2}\v t.
\end{equation}
In order to describe by (\ref{2-9}) a constant velocity motion we have to take
\begin{equation}\label{2-10}
\b=1-\a v^2.
\end{equation}
The functions $\a,\b$ are still arbitrary. However the  field equation (\ref{2-3} has to be satisfied.\\
Calculate  the derivatives
\begin{equation}
\R_t=-\v\b, \qquad \qquad \R_{tt}=0,
\end{equation}
\begin{equation}
\R_x=\e_1-\a v_1\v,  \qquad \qquad \R_{xx}=0.
\end{equation}
Thus
\begin{equation}
f_t=-\frac m{R^3}<\R_t,\R>=\frac {m\b}{R^3}<\v,\R>
\end{equation}
and consequently
\br\label{2-14}
f_{tt}&=&\frac{m\b}{R^3}<\v,\R_t>-3\frac{m\b}{R^5}<\v,\R><\R_t,\R>\nonumber\\
&=&\frac{m\b^2}{R^5}\Big(3<\v,\R>^2-v^2R^2\Big).
\er
As for the spatial part of the  d'Alembertian
\begin{equation}
f_x=-\frac{m}{R^3}<\R_x,\R>,
\end{equation}
\begin{equation}
f_{xx}=-\frac{m}{R^3}<\R_x,\R_x>+3\frac{m}{R^5}<\R_x,\R>^2.
\end{equation}
Thus the Laplacian 
\begin{equation}
\triangle \ f= 3\frac{m}{R^5}\Big(<\R_x,\R>^2+<\R_y,\R>^2+<\R_z,\R>^2\Big)-
\frac{m}{R^3}\Big(R_x^2+R_y^2+R_z^2\Big).
\end{equation}
Using the relation
\begin{equation}
<\R_x,\R>=R_1-\a v_1<v,R>
\end{equation}
we obtain
\begin{equation} 
<\R_x,\R>^2+<\R_y,\R>^2+<\R_z,\R>^2=R^2-2\a<v,R>^2+\a^2v^2<v,R>^2
\end{equation}
and
\begin{equation} 
R_x^2+R_y^2+R_z^2=3-2\a v^2+\a^2v^4.
\end{equation}
Hence the Laplacian takes the form
\br\label{2-21}
\triangle \ f&=& 3\frac{m}{R^5}\Big(R^2-2\a<v,R>^2+\a^2v^2<v,R>^2\Big)
-\frac{m}{R^3}\Big(3-2\a v^2+\a^2v^4\Big)\nonumber\\
&=&3\frac{m}{R^5}<v,R>^2(\a^2v^2-2\a)-\frac{m}{R^3}(\a^2v^2-2\a)v^2\nonumber\\
&=&\frac{m}{R^5}(\a^2v^2-2\a)\Big(3<v,R>^2-v^2R^2\Big)
\er
Since $\square \ f=0$, equations (\ref{2-14}) and (\ref{2-21}) imply that
\begin{equation} 
\b^2=\a^2v^2-2\a.
\end{equation}
By  (\ref{2-14}) the Lorentz parameters $\a$ and $\b$ are 
\begin{equation} 
\b=\frac 1 {\sqrt{1-v^2}}, \qquad\qquad \a=\frac 1{v^2}\Big(1-\frac 1 {\sqrt{1-v^2}}\Big)
\end{equation}
The result is rather obvious. Indeed, beginning with the static 1-particle solution (\ref{2-2}) of the wave equation (\ref{2-3}) and making a Lorentz transformation of coordinates (with opposite velocity) we obtain a solution which describes the inertial motion of the particle.\\ 
Consider now the function
\begin{equation}
f=\frac mR,
\end{equation}
where
\begin{equation}
\R=(\r - \r_0)-\a\dps <\dps,(\r-\r_0)>-\b \p,
\end{equation}
with $\p=\p(t)$ and $\a$ and $\b$ are functions of $|\dps|^2$:
\begin{equation}
\b=\frac 1 {\sqrt{1-|\dps|^2}}, \qquad\qquad \a=\frac 1{|\dps|^2}\Big(1-\frac 1 {\sqrt{1-|\dps|^2}}\Big)
\end{equation}
If $\dps=\overline{v}t$ then (\ref{2-5}) coincides with (\ref{2-5}).\\
Calculate the d'Alembertian of the function $f$. 
\br
\R_t&=&-2\a'\Big(<\dps,\ddps>\dps <\dps,(\r-\r_0)>\Big)-\a\Big(\ddps<\dps,(\r-\r_0)>\Big)\nonumber\\
&&-\a\Big(\dps <\ddps,(\r-\r_0)>\Big)
-2\b'\Big(<\dps,\ddps> \p\Big) -\b \dps 
\er
and
\br
\R_{tt}&=&-4\a''\Big(<\dps,\ddps>^2 <\dps,(\r-\r_0)>\Big)
-2\a'\Big(|\ddps|^2 <\dps,(\r-\r_0)>\Big)\nonumber\\&&
-2\a'\Big(<\dps,\dddps> <\dps,(\r-\r_0)>\Big)
-2\a'\Big(<\dps,\ddps> <\ddps,(\r-\r_0)>\Big)\nonumber\\&&
-4\b''\Big(<\dps,\ddps>^2 \p \Big)-2\b'\Big(|\ddps|^2 \p \Big)
-2\b'\Big(<\dps,\dddps> \p\Big) -2\b'\Big(<\dps,\ddps> \dps\Big) \nonumber\\&&
-2\b'\Big( <\dps,\ddps>\dps \Big)
-\b \ddps. 
\er
Thus
\br
f_t&=&-\frac m{R^3}<\R_t,\R>=-\frac m{R^3}\bigg(
-2\a'\Big(<\dps,\ddps><\dps,\R> <\dps,(\r-\r_0)>\Big)-\nonumber\\&&
\a\Big(<\ddps,\R><\dps,(\r-\r_0)>\Big)
-\a\Big(<\dps,R> <\ddps,(\r-\r_0)>\Big)\nonumber\\
&&-2\b'\Big(<\dps,\ddps> <\p ,\R>\Big)-\b \Big(<\dps ,R>\Big)\bigg)
\er
In the first approximation (for the motion of the particle, which is slow with respect to the speed of the light) we can take
\begin{equation}
\R_t=-\b\dps, \qquad \R_{tt}=-\b\ddps.
\end{equation}
Consequently,
\begin{equation}
f_t=\frac m{R^3}\b <\dps ,R>
\end{equation}
In general, the derivatives of $\a$ and $\b$ contribute terms that are quadratic in $\dps$ and its derivatives and can be neglected.
Thus the second derivative to the same accuracy is 
\br
f_{tt}&=&\frac{m\b}{R^3}\Big(<\ddps ,\R>+<\dps ,\R_t>\Big)
-3\frac{m\b}{R^5}<\dps ,\R><\R_t,\R>
\er
Substitute
\begin{equation}
\R_t=-\b \dps 
\end{equation}
to get
\br
f_{tt}&=&\frac{m\b}{R^3}\Big(
<\ddps ,\R>-\b \ <\dps ,\dps >\Big)
 \ + \ 3\frac{m\b^2}{R^5}<\dps ,\R>^2\nonumber\\
&=&\frac{m\b}{R^3}<\ddps ,\R> \ 
+ \  \ \frac{m\b^2}{R^5}\Big(3<\dps ,\R>^2-R^2|\dps|^2\Big).
\er
As for the spatial derivatives 
\begin{equation}
\R_x=\e_1-\a \dot{\psi}_1\dps,  \qquad \qquad \R_{xx}=0
\end{equation}
\begin{equation}
f_x=-\frac{m}{R^3}<\R_x,\R>
\end{equation}
\begin{equation}
f_{xx}=-\frac{m}{R^3}<\R_x,\R_x> \ + \ 3\frac{m}{R^5}<\R_x,\R>^2
\end{equation}
\begin{equation}
\triangle \ f= 3\frac{m}{R^5}\Big(<\R_x,\R>^2+<\R_y,\R>^2+<\R_z,\R>^2\Big)-
\frac{m}{R^3}\Big(R_x^2+R_y^2+R_z^2\Big)
\end{equation}
Since
\begin{equation}
<\R_x,\R>=R_1-\a\dot{\psi}_1<\dps,R>
\end{equation}
we get
\begin{equation} 
<\R_x,\R>^2+<\R_y,\R>^2+<\R_z,\R>^2=R^2-2\a<\dps,R>^2+\a^2|\dps|^2<\dps,R>^2
\end{equation}
and
\begin{equation} 
R_x^2+R_y^2+R_z^2=3-2\a |\dps|^2+\a^2|\dps|^4
\end{equation}
Thus the Laplacian is 
\begin{equation} 
\triangle f=\frac{m}{R^5}(\a^2|\dps|^2-2\a)\Big(3<\dps,R>^2-|\dps|^2R^2\Big)
\end{equation}
And, for the d'Alembertian
\begin{equation} 
\square f=\frac{m\b}{R^3}<\ddps ,\R>+\frac{m}{R^5}(\b^2+2\a-\a^2|\dps|^2)\Big(3<\dps ,\R>^2-R^2|\dps|^2\Big)
\end{equation}
Using the expressions for the functions $\a,\b$ we obtain 
\begin{equation} 
\square f=\frac{m\b}{R^3}<\ddps ,\R>
\end{equation}
Thus in the case of a linear field equation $\square f=0$ the motion of a one singularity point is inertial i.e.  $\ddps=0$. \\
The reasoning above can be extended to arbitrary number of singularities. Indeed, the equation (\ref{2-3}) is linear so it has a solution  
\begin{equation} 
f=\sum_{i=1}^n\frac {{}^{(i)}m}{{}^{(i)}R},
\end{equation}
where
\begin{equation}
{}^{(i)}\R=(\r - {}^{(i)}\r)-{}^{(i)}\a{}^{(i)}\dps <{}^{(i)}\dps,(\r-{}^{(i)}\r)>-{}^{(i)}\b{}^{(i)} \p,
\end{equation}
with
\begin{equation} 
{}^{(i)}\b=\frac 1 {\sqrt{1-|{}^{(i)}\dps|^2}}, \qquad\qquad 
{}^{(i)}\a=\frac 1{|{}^{(i)}\dps|^2}\Big(1-\frac 1 {\sqrt{1-|{}^{(i)}\dps|^2}}\Big)
\end{equation}
The calculation above yields, approximately
\begin{equation}\label{2-48}
\square f=\sum_{i=1}^n \frac{m_i\b_i}{R_i^3}<\ddps_i,\R_i>.
\end{equation}
Also here, for the field equation is $\square f=0 \ \ $ $f$ is a solution provides that $\ddps_i=0$.\\
Thus, the linear equation describes the free inertial motion of an arbitrary system of singularities.
\sect{Non-linear equations}
In order to describe the interaction between singularities we need to balance the value (\ref{2-48}).
Let us consider  a  new non-linear field equation
\begin{equation} \label{2-49}
\square f=k\eta^{ab}f_{,a}f_{,b}
\end{equation}
where $k$ is a dimensionless constant. It is easy to see that this equation is unique Lorentz-invariant equation which is linear in the second derivatives and quadratic in the first derivatives.\\
The first step is to find  a static spherical-symmetric solution of (\ref{2-49}). Write $f=f(s)$ with $s=x^2+y^2+z^2)$.
So
\begin{equation} 
\triangle f = 2f'+4f''s
\end{equation}
and
\begin{equation} 
\eta^{ab}f_{,a}f_{,b}=-4(f')^2s.
\end{equation}
Thus the equation takes the form
\begin{equation} 
2f''s+3f'=2k(f')^2s
\end{equation}
Inserting $f'=Z$ we obtain
\begin{equation} 
2Z's+3Z-2kZ^2s=0
\end{equation}
Take the new variable 
\begin{equation} 
Z=s^{-3/2}Y
\end{equation}
to obtain
\begin{equation} 
Y'=ks^{-3/2}Y^2.
\end{equation}
Thus 
\begin{equation} 
Y=\frac{\sqrt{s}}{C\sqrt{s}+2k}\qquad ===>\qquad 
f'=\frac 1 {s(C\sqrt{s}+2k)}
\end{equation}
And the unique exact solution of (\ref{2-49}) is
\begin{equation} \label{2-57}
f=-\frac 1k \ln{\Big(1-k\frac{m}{r}\Big)}.
\end{equation}
Note that for the limit $k\to 0$ we obtain $f\to\frac{m}{r}$. The solution (\ref{2-57}) is singular at the point $r=0$. The singularity can also be located in an arbitrary point $\r=\r_0$. \\
The next steep is to consider a moving singularity.\\
We seek for a solution of the form 
\begin{equation} 
f=-\frac 1k \ln{\Big(1-k\frac{m}{R}\Big)}
\end{equation}
for a moving singularity:
\begin{equation} 
\R=(\r - \r_0)-\a\dps <\dps,(\r-\r_0)>-\b \p,
\end{equation}
where $\a$ and $\b$ are functions of $|\dps |^2$.
From the calculations above
\begin{equation} 
\R_t=-\b\dps, \qquad \R_{tt}=-\b\ddps.
\end{equation}
Thus 
\begin{equation} 
f_t=-\frac m{R^3}\frac{<\R,\R_t>}{1-k\frac{m}{R}}=
\frac {m\b}{R^3}\frac{<\R,\dps>}{1-k\frac{m}{R}}.
\end{equation}
And
\br
f_{tt}=\frac{m\b}{R^5}\cdot\frac{3\b<\R,\dps>^2+R^2<\R,\ddps>-\b R^2|\dps|^2}
{1-k\frac{m}{R}}+
\frac{m^2\b^2 k}{R^6}\cdot\frac{<\R,\dps>^2}{(1-k\frac{m}{R})^2}.
\er
As for the spatial derivatives we have
\begin{equation}
\R_x=\e_1-\a\dps<\dps,\e_1>, \qquad \R_{xx}=0
\end{equation}
\begin{equation}
f_x=-\frac m{R^3}\cdot\frac{<\R,\R_x>}{1-k\frac{m}{R}}
\end{equation}
\begin{equation}
f_{xx}=3\frac m{R^5}\cdot\frac{<\R,\R_x>^2}{1-k\frac{m}{R}}-
\frac m{R^3}\cdot\frac{<\R_x,\R_x>}{1-k\frac{m}{R}}+
\frac {km^2}{R^6}\cdot\frac{<\R,\R_x>^2}{(1-k\frac{m}{R})^2}.
\end{equation}
Thus 
\br
\triangle f&=&3\frac m{R^5}\cdot\frac{<\R,\R_x>^2+<\R,\R_y>^2+<\R,\R_z>^2}
{1-k\frac{m}{R}}-\nonumber\\
&&\frac m{R^3}\cdot\frac{<\R_x,\R_x>+<\R_y,\R_y>+<\R_z,\R_z>}{1-k\frac{m}{R}}+\nonumber\\
&&\frac {km^2}{R^6}\cdot\frac{<\R,\R_x>^2+<\R,\R_y>^2+<\R,\R_z>^2}{(1-k\frac{m}{R})^2}
\er
Substitute the value of $\R_x$ to get
\br
\triangle f&=&3\frac m{R^5}\cdot\frac {R^2-2\a<\dps,\R>^2+\a^2|\dps|^2<\dps,\R>^2}{1-k\frac{m}{R}}-
\frac m{R^5}\cdot\frac{3-2\a|\dps|^2 +\a^2|\dps|^4}{1-k\frac{m}{R}}+\nonumber\\&&\frac {km^2}{R^6}\cdot\frac{R^2-2\a<\dps,\R>^2+\a^2|\dps|^2<\dps,\R>^2}{(1-k\frac{m}{R})^2}
\er
Thus the second order l.h.s. of the equation is
\br
\square f&=&\frac{m\b}{R^5}\cdot\frac{3\b<\R,\dps>^2+R^2<\R,\ddps>-\b R^2|\dps|^2}{1-k\frac{m}{R}}+
\frac{m^2\b^2 k}{R^6}\cdot\frac{<\R,\dps>^2}{(1-k\frac{m}{R})^2}\nonumber\\&&
-3\frac m{R^5}\cdot\frac {R^2-2\a<\dps,\R>^2+\a^2|\dps|^2<\dps,\R>^2}{1-k\frac{m}{R}}+\frac m{R^3}\cdot\frac{3-2\a|\dps|^2 +\a^2|\dps|^4}{1-k\frac{m}{R}}-
\nonumber\\&&\frac {km^2}{R^6}\cdot\frac{R^2-2\a<\dps,\R>^2+\a^2|\dps|^2<\dps,\R>^2}{(1-k\frac{m}{R})^2}\nonumber\\
&=&\frac {m\b}{R^3}\cdot\frac{<\R,\ddps>}{1-k\frac{m}{R}}+
\frac m{R^5}(\b^2+2\a-\a^2|\dps|^2)\cdot\frac{(3<\R,\dps>^2-R^2|\dps|^2)}{1-k\frac{m}{R}}+\nonumber\\&&
\frac {m^2k}{R^6}\cdot\frac{-R^2+(\b^2+2\a-\a^2|\dps|^2)<\R,\dps>^2}{(1-k\frac{m}{R})^2}
\er
Using the relation $\b^2=\a^2|\dps|^2-2\a$ we obtain
\begin{equation} 
\square f=\frac {m\b}{R^3}\cdot\frac{<\R,\ddps>}{1-k\frac{m}{R}}-\frac {m^2k}{R^4}\frac 1 {(1-k\frac{m}{R})^2}
\end{equation}
As for the quadratic r.h.s. 
\br
\eta^{ab}f_{,a}f_{,b}&=&\frac {m^2\b^2}{R^6}\cdot\frac{<\R,\dps>^2}{(1-k\frac{m}{R})^2}-
\frac {m^2}{R^6}\cdot\frac{<\R,\R_x>^2+<\R,\R_y>^2+<\R,\R_z>^2}{(1-k\frac{m}{R})^2}\nonumber\\
&=&\frac {m^2\b^2}{R^6}\cdot\frac{<\R,\dps>^2}{(1-k\frac{m}{R})^2}-
\frac {m^2}{R^6}\cdot\frac{R^2-2\a<\dps,\R>^2+\a^2|\dps|^2<\dps,\R>^2}{(1-k\frac{m}{R})^2}\nonumber\\
&=&-\frac {m^2}{R^4}\frac 1 {(1-k\frac{m}{R})^2}+\frac {m^2}{R^6}(\b^2+2\a-\a^2|\dps|^2)\frac{<\R,\dps>^2}{(1-k\frac{m}{R})^2}
\er
Thus 
\begin{equation} 
\eta^{ab}f_{,a}f_{,b}=-\frac {m^2}{R^4}\frac 1 {(1-k\frac{m}{R})^2}
\end{equation}
Thus the field equation (\ref{2-49}) results in 
\begin{equation} 
\frac {m\b}{R^3}\cdot\frac{<\R,\ddps>}{1-k\frac{m}{R}}=0
\end{equation}
This  means $\ddps=0$. So the one point singularity moves with constant velocity.\\
Consider now  a system of $n$ singular points. In this case the ansatz is
\begin{equation} 
f=-\frac 1k \sum^n_{i=1}\ln{\Big(1-k\frac{{}^{(i)}m}{{}^{(i)}R}\Big)}
\end{equation}
where
\begin{equation} 
{}^{(i)}\R=(\r - {}^{(i)}\r)-{}^{(i)}\a\Big({}^{(i)}\dps <{}^{(i)}\dps,(\r-{}^{(i)}\r)>\Big)-{}^{(i)}\b {}^{(i)}\p.
\end{equation}
The linear part of the equation is
\begin{equation} 
\square f=\sum^n_{i=1}\Big( \frac {{}^{(i)}m{}^{(i)}\b}{{}^{(i)}R^3}\cdot\frac{<{}^{(i)}\R,{}^{(i)}\ddps>}{1-k\frac{{}^{(i)}m}{{}^{(i)}R}}-\frac {{}^{(i)}m^2k}{{}^{(i)}R^4}\frac 1 {(1-k\frac{{}^{(i)}m}{{}^{(i)}R})^2}\Big)
\end{equation}
Calculate the nonlinear part
\begin{equation} 
f_t=-\sum^n_{i=1}\Big(\frac {<{}^{(i)}\R_t,{}^{(i)}\R>}{1-k\frac{{}^{(i)}m}{{}^{(i)}R}}\cdot \frac {{}^{(i)}m}{{}^{(i)}R^3}\Big)=
\sum^n_{i=1}\Big(\frac {<{}^{(i)}\dps,{}^{(i)}\R>}{1-k\frac{{}^{(i)}m}{{}^{(i)}R}}\cdot \frac {{}^{(i)}m{}^{(i)}\b}{{}^{(i)}R^3}\Big).
\end{equation}
Thus
\begin{equation} 
(f_t)^2=\sum^n_{i,j=1}\frac {\frac {{}^{(i)}m{}^{(i)}\b}{{}^{(i)}R^3}}{1-k\frac{{}^{(i)}m}{{}^{(i)}R}}\cdot\frac {\frac {{}^{(j)}m{}^{(j)}\b}{{}^{(j)}R^3}}{1-k\frac{{}^{(j)}m}{{}^{(j)}R}}<{}^{(i)}\dps,{}^{(i)}\R><{}^{(j)}\dps,{}^{(j)}\R>.
\end{equation}
As for the spatial derivatives 
\begin{equation} 
f_x=-\sum^n_{i=1}\frac {<{}^{(i)}\R_x,{}^{(i)}\R>}{1-k\frac{{}^{(i)}m}{{}^{(i)}R}}\cdot 
\frac {{}^{(i)}m}{{}^{(i)}R^3}
\end{equation}
\br \label{2-79}
<\nabla f,\nabla f>&=&\sum^n_{i,j=1}\frac {\frac {{}^{(i)}m}{{}^{(i)}R^3}}{1-k\frac{{}^{(i)}m}{{}^{(i)}R}}\cdot\frac {\frac {{}^{(j)}m}{{}^{(j)}R^3}}{1-k\frac{{}^{(j)}m}{{}^{(j)}R}}\Big(<{}^{(i)}\R_x,{}^{(i)}\R><{}^{(j)}\R_x,{}^{(j)}\R>+\nonumber\\
&&<{}^{(i)}\R_y,{}^{(i)}\R><{}^{(j)}\R_y,{}^{(j)}\R>+<{}^{(i)}\R_z,{}^{(i)}\R><{}^{(j)}\R_z,{}^{(j)}\R>\Big)
\er
Using the relation
\begin{equation} 
{}^{(i)}\R_x=\e_1-{}^{(i)}\a{}^{(i)}\dps<{}^{(i)}\dps,\e_1>
\end{equation}
and writing 
\begin{equation} 
<{}^{(i)}\R_x,{}^{(i)}\R>=<\e_1,{}^{(i)}\R>-{}^{(i)}\a<{}^{(i)}\dps,{}^{(i)}\R><{}^{(i)}\dps,\e_1>
\end{equation}
we obtain
\br
&&\Big(<{}^{(i)}\R_x,{}^{(i)}\R><{}^{(j)}\R_x,{}^{(j)}\R>\Big)=
\Big(<\e_1,{}^{(i)}\R><\e_1,{}^{(j)}\R>\Big)-\nonumber\\
&&\qquad\qquad\quad{}^{(j)}\a\Big(<\e_1,{}^{(i)}\R><{}^{(j)}\dps,{}^{(j)}\R><{}^{(j)}\dps,\e_1>\Big)-\nonumber\\
&&\qquad\quad\qquad
{}^{(i)}\a\Big(<\e_1,{}^{(j)}\R><{}^{(i)}\dps,{}^{(i)}\R><{}^{(i)}\dps,\e_1>\Big)+
\nonumber\\
&&\qquad\quad\qquad{}^{(i)}\a{}^{(j)}\a<{}^{(i)}\Big(\dps,{}^{(i)}\R><{}^{(i)}\dps,\e_1><{}^{(j)}\dps,{}^{(j)}\R><{}^{(j)}\dps,\e_1>\Big).
\er
Thus the brackets in (\ref{2-79}) are
\br
\Big({\quad }\Big)&=&<{}^{(i)}\R,{}^{(j)}\R>-{}^{(j)}\a\Big(<{}^{(j)}\dps,{}^{(i)}\R><{}^{(j)}\dps,{}^{(j)}\R>\Big)-\nonumber\\
&&{}^{(i)}\a\Big(<{}^{(i)}\dps,{}^{(j)}\R><{}^{(i)}\dps,{}^{(i)}\R>\Big)+\nonumber\\
&&{}^{(i)}\a{}^{(j)}\a\Big(<{}^{(i)}\dps,{}^{(i)}\R><{}^{(i)}\dps,{}^{(j)}\dps><{}^{(j)}\dps,{}^{(j)}\R>\Big).
\er
Thus, the r.h.s. of the field equation is
\brn
k\eta^{ab}f_{,a}f_{,b}&=&-\sum^n_{i,j=1}\frac {\frac {{}^{(i)}m}{{}^{(i)}R^3}}{1-k\frac{{}^{(i)}m}{{}^{(i)}R}}\cdot\frac {\frac {{}^{(j)}m}{{}^{(j)}R^3}}{1-k\frac{{}^{(j)}m}{{}^{(j)}R}}\Big[<{}^{(i)}\R,{}^{(j)}\R>+\\
&&<{}^{(i)}\dps,{}^{(j)}\R><{}^{(i)}\dps,{}^{(i)}\R> \Big({}^{(i)}\b{}^{(j)}\b+
{}^{(i)}\a+{}^{(j)}\a-{}^{(i)}\a{}^{(j)}\a<{}^{(i)}\dps,{}^{(j)}\dps>\Big)\Big]
\ern
Extracting in this expression the nonlinear part we obtain
\brn
k\eta^{ab}f_{,a}f_{,b}&=&-k\sum_i\frac {{}^{(i)}m^2k}{{}^{(i)}R^4}\frac 1 {(1-k\frac{{}^{(i)}m}{{}^{(i)}R})^2}
-\sum_{i\ne j}\frac {\frac {{}^{(i)}m}{{}^{(i)}R^3}}{1-k\frac{{}^{(i)}m}{{}^{(i)}R}}\cdot\frac {\frac {{}^{(j)}m}{{}^{(j)}R^3}}{1-k\frac{{}^{(j)}m}{{}^{(j)}R}}\Big[<{}^{(i)}\R,{}^{(j)}\R>+\\
&&<{}^{(i)}\dps,{}^{(j)}\R><{}^{(i)}\dps,{}^{(i)}\R> \Big({}^{(i)}\b{}^{(j)}\b+
{}^{(i)}\a+{}^{(j)}\a-{}^{(i)}\a{}^{(j)}\a<{}^{(i)}\dps,{}^{(j)}\dps>\Big)\Big]
\ern
Thus the field equation takes the form
\br\label{2-85}
&&\sum^n_{i=1} \frac {{}^{(i)}m{}^{(i)}\b}{{}^{(i)}R^3}\cdot\frac{<{}^{(i)}\R,{}^{(i)}\ddps>}{1-k\frac{{}^{(i)}m}{{}^{(i)}R}}=-k\sum_{i\ne j}\frac {\frac {{}^{(i)}m}{{}^{(i)}R^3}}{1-k\frac{{}^{(i)}m}{{}^{(i)}R}}\cdot\frac {\frac {{}^{(j)}m}{{}^{(j)}R^3}}{1-k\frac{{}^{(j)}m}{{}^{(j)}R}}\bigg(<{}^{(i)}\R,{}^{(j)}\R>+\nonumber\\
&&<{}^{(i)}\dps,{}^{(j)}\R><{}^{(i)}\dps,{}^{(i)}\R> \Big({}^{(i)}\b{}^{(j)}\b+
{}^{(i)}\a+{}^{(j)}\a-{}^{(i)}\a{}^{(j)}\a<{}^{(i)}\dps,{}^{(j)}\dps>\Big)\bigg)
\er
For the approximation of  slow motions it is
\br\label{2-86}
&&\sum^n_{i=1} \frac {{}^{(i)}m}{{}^{(i)}R^3}\cdot\frac{<{}^{(i)}\R,{}^{(i)}\ddps>}{1-k\frac{{}^{(i)}m}{{}^{(i)}R}}=-k\sum_{i\ne j}\frac {\frac {{}^{(i)}m}{{}^{(i)}R^3}}{1-k\frac{{}^{(i)}m}{{}^{(i)}R}}\cdot\frac {\frac {{}^{(j)}m}{{}^{(j)}R^3}}{1-k\frac{{}^{(j)}m}{{}^{(j)}R}}<{}^{(i)}\R,{}^{(j)}\R>
\er
The two sides of this equation are functions of an arbitrary point $x$. Choose the singularity point $i=p$ and consider the arbitrary point $x$ to be close to this singularity. It follows that
\begin{equation}\label{2-87}
{}^{(p)}\R\to 0 \text{\quad and\quad } {}^{(i)}\R\to \bar{R}_{ip}\text{\quad for\quad }
i\ne k,
\end{equation}
where $\bar{R}_{ip}$ is a vector from the point $i$ to the point $p$.\\ 
In the l.h.s. of the equation (\ref{2-86}) there is one singular term
\begin{equation}\label{si1}
\frac{{}^{(p)}m}{{}^{(p)}R^3}\frac{<{}^{(p)}\R,{}^{(p)}\ddps>}{1-k\frac{{}^{(p)}m}{{}^{(p)}R}}
\end{equation}
The singular term in the r.h.s. of (\ref{2-86}) is
\begin{equation}\label{si2}
-k\frac {\frac {{}^{(p)}m}{{}^{(p)}R^3}}{1-k\frac{{}^{(p)}m}{{}^{(p)}R}}\sum_{j\ne p}\frac {\frac {{}^{(j)}m}{{}^{(j)}R^3}}{1-k\frac{{}^{(j)}m}{{}^{(j)}R}}<{}^{(p)}\R,{}^{(j)}\R>
\end{equation}
The terms (\ref{si1}) and (\ref{si2}) are $O(R^{-2}$ near the singularity. When these are inserted to the r.h.s and l.h.s. of (\ref{2-86}), respectively, the remainder will be $O(R^{-1}$ only if: 
\begin{equation}\label{2-89}
<{}^{(p)}\R,{}^{(p)}\ddps>=-k\sum_{j\ne p}\frac {\frac {{}^{(j)}m}{{}^{(j)}R^3}}{1-k\frac{{}^{(j)}m}{{}^{(j)}R}}<{}^{(p)}\R,{}^{(j)}\R>
\end{equation}
This way, the strength of the singularity is diminished.\\
Take into account that  the point $x$ is still arbitrary. Hence (\ref{2-89}) can be valid only if 
\begin{equation}
{}^{(p)}\ddps=-k\sum_{j\ne p}\frac {\frac {{}^{(j)}m}{{}^{(j)}R^3}}{1-k\frac{{}^{(j)}m}{{}^{(j)}R}}{}^{(j)}\R
\end{equation}
For the limiting values in (\ref{2-87}) 
\begin{equation}
{}^{(p)}\ddps=-k\sum_{j\ne p}\frac {{}^{(j)}m\R_{jp}}{R_{jp}^3}\frac 1 {1-k\frac{{}^{(j)}m}{R_{jp}}}
\end{equation}
The second fraction differs from 1, significantly, only for small distances comparable to the Schwarzschild radius $r=km$. Thus, we neglect it. \\
It follows that
 \begin{equation}\label{2-92}
{}^{(p)}\ddps=-k\sum_{j\ne p}\frac {{}^{(j)}m\R_{jp}}{R_{jp}^3}
\end{equation}
For a system of two singularity points  
\begin{equation}\label{2-93}
{}^{(1)}\ddps=-k\frac {{}^{(2)}m\R_{21}}{R_{21}^3}
\end{equation}
If $k<0$ then (\ref{2-92})  and (\ref{2-93}) result in attraction between the particles. The absolute value of $k$ is unimportant, since it amounts to rescaling of the mass.\\
This way the Newton law
\begin{equation}
{}^{(p)}\ddps=\sum_{j\ne p}\frac {{}^{(j)}m\R_{jp}}{R_{jp}^3}.
\end{equation}
is obtained.


\end{document}